\documentclass[%
 reprint,
 amsmath,amssymb,
 aps,
]{revtex4-2}

\usepackage{graphicx}
\usepackage{xcolor}
\usepackage{dcolumn}
\usepackage{bm}
\usepackage{hyperref}

\begin{document}

\preprint{APS/123-QED}

\title{Unlocking Photon–Magnon Interplay via Saturation Magnetization}

\author{Sachin Verma$^{1}$}
\email{sachinverma.rs.phy22@iitbhu.ac.in}
\author{Jiten Mahalik$^{2}$}
\author{Abhishek Maurya$^{1}$}
\author{Rajeev Singh$^{1}$}
\author{Biswanath Bhoi$^{1}$}
 \email{Corresponding author, E-mail: biswanath.phy@itbhu.ac.in}
\affiliation{$^{1}$Nano-Magnetism and Quantum Technology Lab,
 Department of Physics, Indian Institute of Technology (Banaras Hindu University) Varanasi, Varanasi - 221005, India.}

\affiliation{$^{2}$National Institute of Science Education and Research, Bhubaneswar, Odisha, India}%




\date{\today}

\begin{abstract}
Photon–magnon hybrid systems present a promising platform for the development of next-generation devices in quantum information processing and quantum sensing technologies. In this study, we investigate the control of photon–magnon coupling (PMC) strength through systematic variation of the saturation magnetization $M_{\mathrm{s}}$ in a planar hexagonal-ring resonator (HRR) integrated with a yttrium iron garnet (YIG) thin film configuration. Using full-wave numerical simulations in CST Microwave Studio, we demonstrate that tuning the $M_{\mathrm{s}}$ of the YIG film from 1750 Oe to 900 Oe enables systematic control over the coupling strength across the 127–51 MHz range at room temperature. To explain the observed PMC dynamics, we develop a semiclassical analytical model based on electromagnetic theory, that accurately reproduces the observed coupling behavior, revealing the key role of spin density in mediating the light–matter interaction. The model is further extended to include the effects of variable magnon damping across different $M_{\mathrm{s}}$ values, enabling broader frequency control. These findings establish $M_{\mathrm{s}}$ as a key tuning parameter for tailoring PMC, with direct implications for the design of tunable hybrid systems for reconfigurable quantum devices.
\end{abstract}

\maketitle


\section{\label{sec:level1}Introduction}
Hybrid Quantum Systems (HQS) have emerged as a versatile platform where distinct physical excitations such as photons, magnons, phonons, plasmons, electrons, and even thermal quasiparticles are coupled to harness their complementary advantages while overcoming the limitations inherent to individual subsystems \cite{zhang2023review, li2020hybrid, xiang2013hybrid, kurizki2015quantum, lachance2019hybrid,yuan2022quantum}. By enabling controlled interactions among these diverse quantum degrees of freedom, HQS provides a powerful framework for developing next-generation quantum technologies. These include high-precision quantum sensors \cite{kantsepolsky2023exploring,degen2017quantum}, coherent quantum transducers \cite{wang2022quantum} for interfacing different frequency domains, robust quantum memories \cite{heshami2016quantum}, and scalable architectures for quantum information processing \cite{nielsen2010quantum} and communication. 

Among the various types of HQS, photon–magnon coupled (PMC) platforms where microwave photons coherently interact with magnons have attracted significant attention \cite{li2020hybrid}. By combining the long coherence times and magnetic tunability of magnons with the high-speed, low-loss transmission capabilities of photons, these systems offer a promising route toward quantum signal processing, memory, and transduction applications \cite{lachance2019hybrid}. In recent years, photon–magnon coupling (PMC) systems have achieved remarkable progress \cite{li2020hybrid}, extending beyond strong coupling and coherent information exchange to enable the observation of exotic phenomena such as level attraction \cite{yang2019control}, nonreciprocal behavior \cite{wang2019nonreciprocity}, and various aspects of non-Hermitian physics \cite{el2018non,ashida2020non}, including exceptional points \cite{miri2019exceptional,ozdemir2019parity} and, more recently, exceptional surfaces \cite{cerjan2019whole}. Realizing a broad range of such exotic phenomena in HQS critically depends on the ability to precisely control PMC. Several strategies have been developed to modulate the coupling strength, such as engineering the geometry \cite{wagle2024controlling} and mode volume of microwave resonators \cite{mckenzie2019low}, and dynamically tuning the relative detuning between photon and magnon modes through external magnetic fields \cite{rao2023meterscale}. Furthermore, the spatial positioning and orientation of the magnetic element within the resonator’s field profile provide additional fine-tuning of the coupling rate \cite{maurya2024room,verma2024control}. These control techniques underscore the pivotal role of PMC tunability in advancing both the fundamental understanding and technological applications of hybrid quantum systems.

Despite significant progress in controlling PMC strength, existing methods face inherent limitations. Geometric modifications of resonators and spatial repositioning of magnetic elements offer only static or limited tunability, often requiring complex fabrication or mechanical adjustments. Frequency detuning through external magnetic fields, while widely used, affects both the magnon and photon modes simultaneously, making it difficult to isolate the effect on coupling strength alone. Moreover, dynamic modulation approaches involving parametric drives or nonlinear interactions introduce added complexity and may compromise coherence. In this context, one often overlooked yet fundamentally promising avenue is the direct tuning of the saturation magnetization $M_{\mathrm{s}}$ of the magnetic material itself. Since the coupling strength scales with the square root of spin density \cite{li2020hybrid,lachance2019hybrid}, altering $M_{\mathrm{s}}$ directly modifies the number of spins available for interaction, providing a clean and intrinsic means of tuning PMC without perturbing resonator geometry or mode structure. 

Motivated by these considerations, we propose and investigate a strategy centered on magnetic saturation tuning as a direct and versatile control knob for PMC. In this work, we explore a novel approach for controlling the strength of PMC by systematically varying the $M_{\mathrm{s}}$ of the magnetic system, which excites magnons to hybridize with photons generated by a microwave resonator. Using full-wave numerical simulations in CST Microwave Studio, we demonstrate that tuning $M_{\mathrm{s}}$ enables a wide and systematic control of the coupling strength range at room temperature. To further understand the PMC dynamics, we develop a semiclassical analytical model based on electromagnetic theory, which accurately reproduces the observed behavior and highlights the key role of spin density in mediating light–matter interactions. By incorporating magnon damping variations with $M_{\mathrm{s}}$, the model extends the tunability across a wider frequency range. These findings establish $M_{\mathrm{s}}$ as an effective tuning parameter for engineering PMC, offering new opportunities for the development of reconfigurable hybrid quantum systems and devices.

\section{\label{sec:level2}Numerical Modeling and Simulation Setup}
Figure 1 presents the schematic of the photon–magnon hybrid system, illustrating the simulation setup comprising a YIG thin film coupled to a hexagonal-ring resonator (HRR) via a microstrip transmission line in a planar geometry. In this configuration, the HRR supports the photon mode, while the YIG film serves as the medium for magnon excitation. To investigate their interaction, full-wave electromagnetic simulations were performed using CST Microwave Studio. 

\begin{figure}[htbp]
    \centering
    \includegraphics[width=\linewidth]{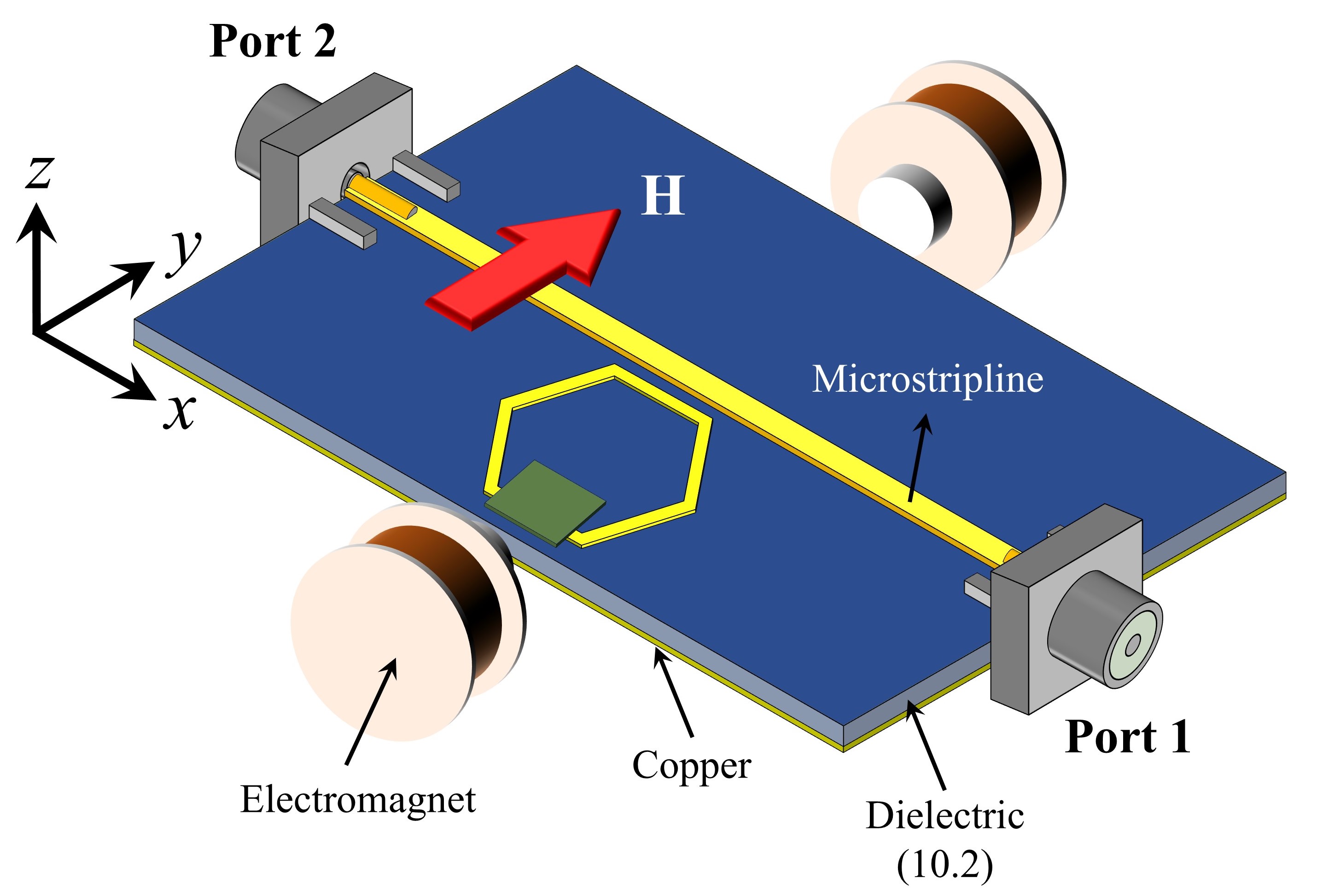}
    \caption{Schematic representation of the simulation setup for a magnon–photon hybrid system comprising a YIG film (green), a microstrip line (dark yellow), a dielectric material (navy blue), and an HRR (dark yellow). The YIG film is positioned at the edge of the HRR, which is placed adjacent to the microstrip line. An external magnetic field (H) is applied uniformly in-plane along the y-axis.}
    \label{F1}
\end{figure}

The microstrip transmission line was placed on the front side of the sample to efficiently excite both the HRR and the YIG film, with a continuous ground plane positioned on the back side. When an alternating current at microwave frequencies is driven through the microstrip, it generates a localized microwave magnetic field around the microstripline and an electric field that terminates perpendicularly onto the ground plane. The microwave magnetic field excites the HRR, which acts as a parallel LC resonator, exibiting a quasi-static photon mode at its resonant frequency. To enable magnon excitation, a uniform in-plane external bias magnetic field $H$ was applied across the YIG film, tuning the magnon resonance frequency by aligning the magnetization and enabling coherent and uniform spin precession at the desired frequency. YIG was selected due to its high spin density and exceptionally low magnetic damping, making it an ideal material for coherent information processing applications. The detailed dimensions of the HRR, microstrip line, and YIG film, along with additional simulation parameters, are provided in Ref. \cite{verma2024control}.

\section{\label{sec:level1}Results and discussion }
To investigate dynamic interaction between photon and magnon, the transmission characteristics $\lvert S_{21} \rvert$ was analyzed by systematically varying the strength of $H$ applied along the y-axis. This method allowed for the examination of both the individual responses of the HRR and the YIG film, as well as their collective behavior in the hybrid configuration. The photon mode resonance frequency of the HRR was identified at $\frac{\omega_c}{2\pi}$ = 5.33 GHz and remained invariant with respect to changes in the applied magnetic field. This field-independence arises because the resonance frequency, given by $\omega_c = 1/LC$, is determined solely by the resonator’s inductance and capacitance, and is unaffected by external magnetic fields. The damping factor of the HRR was calculated as $\beta = \frac{\Delta\omega_c}{\omega_c} = 4.69 \times 10^{-3}$, where $\Delta\omega_c$ denotes the half-width at half-maximum of the $\lvert S_{21} \rvert$ response. These parameters were extracted from numerical simulations conducted on the HRR structure alone, without the inclusion of the YIG film. In contrast, the resonance frequency of the YIG film displayed a strong dependence on the applied magnetic field, shifting to higher frequencies as the field strength increased. This behavior, observed through simulations performed solely on the YIG film, follows the Kittel relation: $\omega_r = \gamma \sqrt{H(H + M_{\mathrm{s}})}$, where $\gamma = 1.76 \times 10^{7}$ rad/Oe.s is the gyromagnetic ratio and $M_{\mathrm{s}}$ is the saturation magnetization (= 175 mT for YIG).

In the HRR–YIG hybrid system, two distinct resonance peaks were observed. The photon mode peak associated with the HRR remained nearly constant throughout the H sweep, while the magnon mode peak corresponding to the YIG film shifted progressively to higher frequencies with increasing magnetic field. As the magnon mode approached the photon mode frequency, an increase in its amplitude was observed. After crossing the photon mode, the amplitude of the magnon mode gradually decreased. (See Fig. S1 in supplementary file) This characteristic evolution indicates strong coupling between the photon mode of the HRR and the magnon mode of the YIG film.  



\subsection{\label{sec:citeref} Effect of Magnetization on Photon-Magnon Interactions}
To investigate the influence of $M_{\mathrm{s}}$ on the photon–magnon interaction and its effect on coupling strength in the HRR–YIG hybrid system, we conducted a series of simulations for various $M_{\mathrm{s}}$ values of the YIG thin film, specifically 175 mT, 150 mT, 120 mT, and 90 mT. All other material and geometrical parameters of the YIG film and HRR were kept constant. The material properties of the YIG film used in these simulations were: gyromagnetic ratio $\gamma = 1.76 \times 10^{7}$ rad/Oe.s, Gilbert damping constant $\alpha = 1.4 \times 10^{-5}$, and magnetic anisotropy constant $K_{\mathrm{a}} = 0$. Figures 2(a) – (d) present the transmission spectra $\lvert S_{21} \rvert$ as a function of $f$ and $H$ for the different $M_{\mathrm{s}}$ values. The colormap background in each plot represents the simulation data, illustrating the photon–magnon hybridization in the system.

We begin the discussion of the simulation results with the case of $M_{\mathrm{s}}$ = 175 mT, as shown in Fig. 2(a). In this case, when the magnon mode frequency is tuned to resonance with the microwave photon mode, a characteristic normal-mode anti-crossing is observed. This phenomenon, indicative of coherent coupling between the photon and magnon modes, occurs at the coupling center, located at $H_{\mathrm{cent}}$ = 121 mT. The normal-mode anti-crossing behavior, indicating strong coupling, is similarly observed for the other $M_{\mathrm{s}}$ values of 150 mT, 120 mT, and 90 mT, as seen in Figs. 2(b) – (d). However, a clear trend emerges upon examining the coupling centers across the different $M_{\mathrm{s}}$ values. Specifically, the coupling centers $H_{\mathrm{cent}}$ for $M_{\mathrm{s}}$ = 150 mT, 120 mT, and 90 mT shift to higher magnetic fields, with values of 128.5 mT, 138.5 mT, and 150 mT, respectively. On the other hand, the splitting between the hybridized modes is observed to decrease monotonically as the $M_{\mathrm{s}}$ decreases from 175 mT to 90 mT. This shift of the coupling center to higher magnetic fields, accompanied by a reduction in mode splitting, is a direct consequence of the decrease in $M_{\mathrm{s}}$, further emphasizing the tunability of the photon–magnon interaction through variations in the saturation magnetization. To emphasize this trend, vertical white lines are used to mark the coupling centers for $M_{\mathrm{s}}$ = 175 mT and $M_{\mathrm{s}}$ = 90 mT, clearly illustrating the total shift of 29 mT in the magnetic field. These results suggest that the strength and characteristics of the PMC in the hybrid system can be systematically controlled by varying the $M_{\mathrm{s}}$ of the magnetic system, providing a valuable mechanism for tuning the interaction in potential quantum devices. 

To gain deeper insight into the observed behavior of the photon–magnon coupling and to quantitatively describe the dependence of the hybridized mode characteristics on the saturation magnetization, a theoretical model is essential. By establishing a suitable theoretical framework, we can better understand the underlying physical mechanisms governing the coupling strength, observed shifts in coupling center, and mode splitting. In the following section, we present the theoretical modeling of the HRR–YIG hybrid system, which captures the essential features observed in the simulations and provides a quantitative basis for further analysis.
\begin{figure}[htbp]
    \centering
    \includegraphics[width=\linewidth]{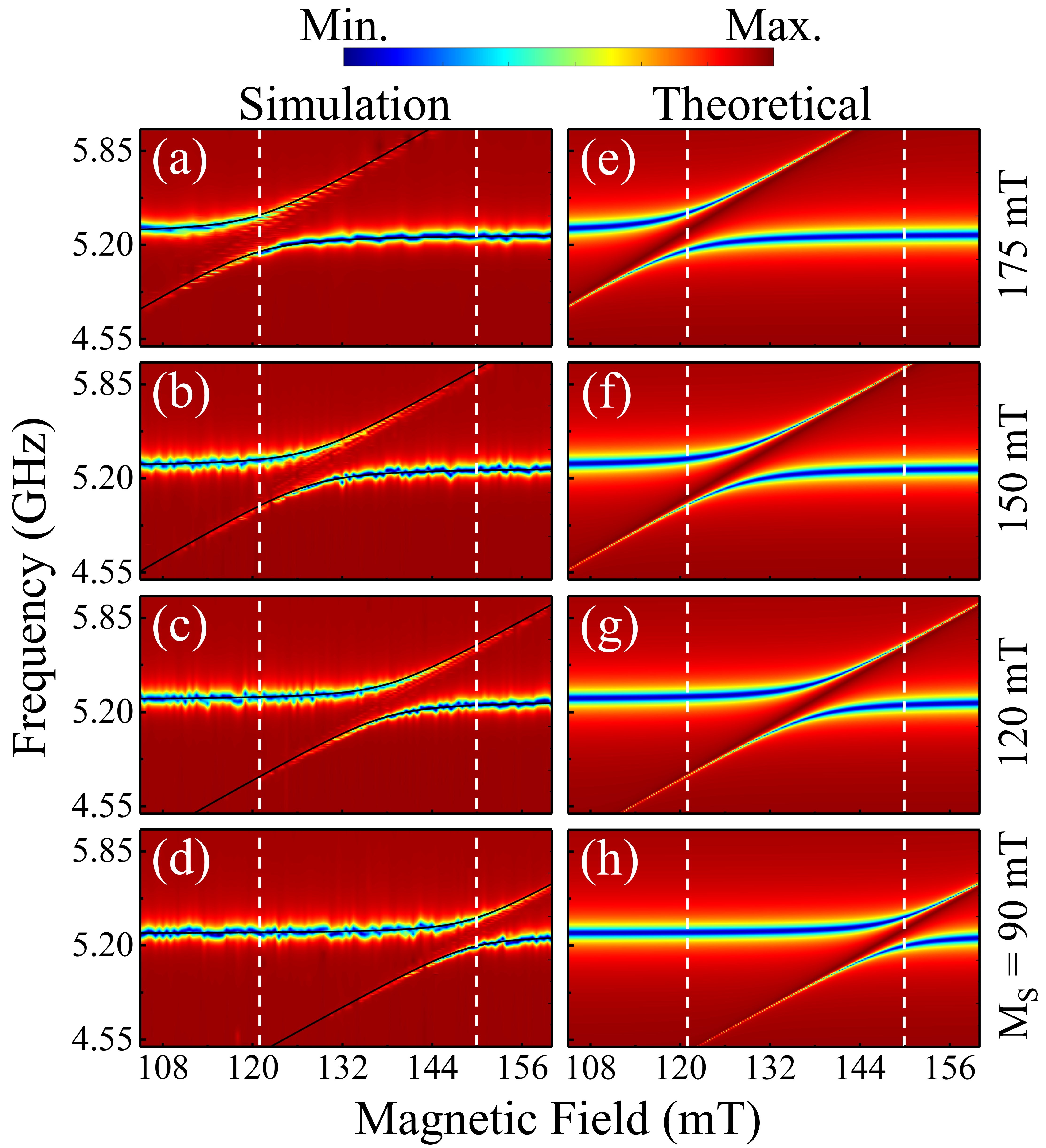}
    \caption{Scanned $\left[f, H\right]$ dispersion of microwave $\lvert S_{21} \rvert$ power plotted for different saturation magnetization $\left(M_{\mathrm{s}}\right)$ values: (a–d) simulation results, and (e–h) theoretical calculations based on an electrodynamic model. The white lines indicate the shift of the coupling center as $M_{\mathrm{s}}$ decreases from 175 mT to 90 mT.}
    \label{F2}
\end{figure}

\vspace{-0.5cm}
\subsection{\label{sec:citeref} Theoretical Formalism}
The photon-magnon interactions in hybrid systems are effectively described by a semiclassical framework that combines the microwave LCR circuit model with the Landau-Lifshitz-Gilbert (LLG) dynamics. These interactions originate from two classical coupling mechanisms. The first, Faraday induction \cite{silva1999inductive}, induces a voltage in the LCR circuit due to the precession of the magnetization. The second, based on Ampère’s law \cite{bai2015spin}, involves magnetic fields generated by the circuit's current, which in turn affect the magnetization. Let’s consider a microwave current $j_0^+ = j_0 e^{-i\omega t}$ flowing through the microstrip line. According to Ampere’s law, this current generates a microwave magnetic field $h_{line} = h e^{-i\omega t}$ around the microstripline, characterized by an amplitude $|h|$ and an angular frequency $\omega$, and the magnetization corresponding to the magnon is given by $m^+ = m e^{-i\omega t}$.

The magnetization precession in the YIG film is described by the Landau-Lifshitz-Gilbert (LLG) equation\cite{bhoi2019abnormal,grigoryan2019cavity},
\begin{eqnarray}
\frac{d\mathbf{m}}{dt} = -\gamma\, \mathbf{m} \times \mathbf{H}_{\text{eff}} + \alpha\, \mathbf{m} \times \frac{d\mathbf{m}}{dt}
\label{eq:one}
\end{eqnarray}
where $\mathbf{m} = \mathbf{M}/\mathbf{M_s}$ represents the magnetization vector, with gyromagnetic ratio $\gamma/2\pi$, intrinsic Gilbert damping parameter $\alpha$, and saturation magnetization $\mathbf{M_{\mathrm{s}}}$ of the YIG film. $\mathbf{H}_{\text{eff}}$ is the effective magnetic field on the YIG film as $\mathbf{H}_{\text{eff}} = H_{\text{dc}} + \mathbf{h}_{\text{line}} + \mathbf{h}_{\text{HRR}} = H\hat{\mathbf{i}} + \mathbf{h} e^{-i\omega t}$, where $H_{\text{dc}}$ is the externally applied DC magnetic field along the y-axis. The magnetizations in the YIG are influenced by the effective field $\mathbf{h} e^{-i\omega t}$, which is the sum of two time-dependent magnetic fields $\mathbf{h}_{\text{line}}$ from the feeding line and $\mathbf{h}_{\text{HRR}}$ due to the $\mathbf{j}^+$. Using a linearized form of the magnetization direction, the magnetization variation is given as, $\mathbf{m} = \mathbf{M_{\mathrm{s}}} \hat{\mathbf{i}} + \mathbf{m}^+$, where $\mathbf{m}^+$ is the oscillating component of the magnetization and $\mathbf{M_{\mathrm{s}}}$ is the magnetization along the z direction. Assuming $\mathbf{m}^+ \ll \mathbf{M_{\mathrm{s}}}$, the LLG equation can be simplified in the rotational frame \cite{bai2015spin,harder2016study} as,
\begin{eqnarray}
\left( \omega - \omega_r + i \alpha \omega \right) m^{+}(t) - i \omega_m K_A j^{+}(t) = 0
\label{eq:two}
\end{eqnarray}
where $m^{+} = m_x + i m_y$ , represents the in-plane magnetization for the magnon mode, $\omega_m=\gamma M_{\mathrm{s}}$ \cite{harder2016study,grigoryan2018synchronized}. The ferromagnetic resonance (FMR) frequency is given by $\omega_r = \gamma \sqrt{H(H + M_s)}$, and $j^{+} = i (h^{+})_{\mathrm{HRR}} K_A$, is the net microwave current in the HRR, and $K_A$ is the coupling parameter that determines the phase relation between the HRR’s photon and YIG’s magnon modes due to Ampere’s law.

As there is a microwave current $j_0^+$ flowing through the microstrip line, it induces a voltage of $V_1=L \frac{d j_0^{+}}{dt}$ in the HRR, and due to the presession of spins of YIG, an additional driving voltage $V_2=i K_F L \frac{d m^{+}}{dt}$, is also induced in the HRR according to the Faraday induction law. Where $K_{\mathrm{F}}$ is the coupling parameter to account for the phase relation between the HRR's photon mode and YIG's magnon modes. This induced voltage $V^{+} = L \frac{d j_0^{+}}{dt} + i K_F L \frac{d m^{+}}{dt}$, generates a microwave current in the HRR, as expressed by $V^{+} = Z_{\mathrm{HRR}} j^{+}$, where $Z_{\mathrm{HRR}}$ being the resonator’s impedance and $j^{+}$ is the net current in the HRR. Now, according to an equivalent LCR circuit model \cite{bloembergen1962phys,grigoryan2018synchronized,grigoryan2019thermally},
\begin{eqnarray}
V^{+} = R j^{+} + \frac{1}{C} \int j^{+} \, dt + L \frac{d j^{+}}{dt}
\label{eq:three}
\end{eqnarray}
where, R, C, and L represent the resistance, capacitance, and inductance, respectively. By solving the above equation, we get,
\begin{eqnarray}
\left( \omega^2 - {\omega_c}^2+ 2i \beta \omega {\omega_c}\right) j^{+}(t) + i {\omega}^2 K_F m^{+}(t) \nonumber \\
= {\omega}^2 {j_0}^{+}
\label{eq:four}
\end{eqnarray}
where, $\omega_{\mathrm{c}}=1/\sqrt{LC}$ is the angular resonance frequency of the HRR with $\beta = R/\left( 2L\omega_c \right) $, where $\beta$ is the damping parameter with resistance $R$.  To simultaneously solve Eqs. (2) and (4), the matrix form is reformulated as:
\begin{equation}
\begin{bmatrix} \omega^2 - {\omega_c}^2+ 2i \beta \omega {\omega_c} & i {\omega}^2 K_F \\ - i \omega_m K_A & \omega - \omega_r + i \alpha \omega \end{bmatrix} \begin{bmatrix} j^{+} \\ m^{+} \end{bmatrix} = \begin{bmatrix} {\omega}^2 {j_0}^{+} \\ 0 \end{bmatrix}
\label{eq:five}
\end{equation}
Let, 
\begin{equation}
\Omega = \begin{bmatrix} \omega^2 - {\omega_c}^2+ 2i \beta \omega {\omega_c} & i {\omega}^2 K_F \\ - i \omega_m K_A & \omega - \omega_r + i \alpha \omega \end{bmatrix} 
\label{eq:six}
\end{equation}
Now, equation (5) can be written as,
\begin{equation}
\Omega \begin{bmatrix} j^{+} \\ m^{+} \end{bmatrix} = \begin{bmatrix} {\omega}^2 {j_0}^{+} \\ 0 \end{bmatrix}
\label{eq:seven}
\end{equation}

The first row of the matrix in Eq. (5) represents the coupling of the magnetization dynamics in the YIG film to the currents in the resonator, describing the LCR circuit influenced by the motion of magnetizations. Conversely, the second row of the matrix captures the influence of the net resonator currents on the magnetization dynamics within the YIG film. The solution of $\det(\Omega)=0$ yields the complex frequency, where its real part corresponds to the dispersion spectrum, and its imaginary part reflects the dissipation in HRR-YIG hybrid system. To understand the observed dispersion spectra, we solve $\det(\Omega)=0$, which gives:
\begin{eqnarray}
\left( \omega^2 - {\omega_c}^2+ 2i \beta \omega {\omega_c} \right) \left( \omega - \omega_r + i \alpha \omega \right) \nonumber \\
-\left( {\omega}^2 K_F \right) \left( \omega_m K_A \right) = 0
\label{eq:four}
\end{eqnarray}
under the resonance condition, where $\omega_c =  \omega_r$, above equation can be written as, 
\begin{equation}
\tilde{\omega}_\pm = \frac{1}{2} \left[ \left( \tilde{\omega}_r + \tilde{\omega}_c \right) \pm \sqrt{\left(\tilde{\omega}_r - \tilde{\omega}_c\right)^2 + 2K^2\omega_m \omega_c}\right]
\label{eq:nine}
\end{equation}
here, $\tilde\omega_c = \omega_c - i\beta \omega_c$, and $\tilde\omega_r = \omega_r - i\alpha \omega_r$, where $\beta$ and $\alpha$ denote the intrinsic damping rates of the photon and magnon modes, respectively. We have calculated the complex eigenvalues of two coupled modes expressed as $\tilde \omega_{\pm} = \omega_{\pm} - i \Gamma_{\pm}$ \cite{harder2016study,zhang2014strongly,bai2015spin}. Here the $\omega_{\pm}$  represents the real part, corresponding to the higher and lower energy modes in the dispersion curve, while $\Gamma_{\pm}$ is the imaginary part, which describes the linewidth evolution of the coupled modes. The frequency gap between the two modes at the anti-crossing center is given by $\left( \omega_{+} - \omega_{-} \right)/2\pi$, which can be determined from Eq. (9). The net coupling strength is defined as $\Delta = \left( \omega_{+} - \omega_{-} \right)/4\pi$ representing half of the frequency gap between the upper and lower branches at the resonance point where $\omega_c =  \omega_r$. So, $\Delta$ can be expressed as,
\begin{equation}
    \Delta = \frac{\omega_c}{4\pi} \sqrt{\frac{2K^2\gamma M_{\mathrm{s}}}{\omega_c}-\left( \beta - \alpha\right)^2}
    \label{eq:ten}
\end{equation}

To determine the coupling strengths, Eqs. (9) and (10) were fitted to the lower- and higher-frequency branches as indicated by the black solid lines in Figs. 2(a)–2(d). The simulation results are in excellent agreement with the fitted curves, confirming the accuracy of the theoretical model. The resulting values of $\Delta$ were 127.6 MHz, 121.4 MHz, 113 MHz, and 97.68 MHz, corresponding to saturation magnetizations $M_{\mathrm{s}}$ values of 175 mT, 150 mT, 120 mT, and 90 mT, respectively. This decreasing trend in coupling strength with decreasing $M_{\mathrm{s}}$ can be understood by considering the fundamental dependence of photon–magnon coupling on the magnetization of the system. In magnetic materials like YIG, when all net spins precess coherently in phase, the collective excitation known as the Kittel mode couples strongly to the HRR cavity mode. In this regime, the coupling strength $\Delta$ scales with the square root of the total number of spins involved, i.e., $\Delta \propto \sqrt{N}$ \cite{lachance2019hybrid,li2020hybrid}. The saturation magnetization $M_{\mathrm{s}}$, which represents the maximum magnetization when all spins are aligned, is directly proportional to the spin density via the relation $M_{\mathrm{s}} \propto N g_{\mathrm{s}}\mu_{\mathrm{B}}/V$, where $N/V$ is the spin density (number of spins per unit volume), $g_{\mathrm{s}}$ is the Lande g-factor, and $\mu_{\mathrm{B}}$ is the Bohr magneton \cite{kittel2018introduction,blundell2001magnetism}. Combining these relations yields $\Delta \propto \sqrt{M_{\mathrm{s}}}$, indicating that the coupling strength is fundamentally linked to the square root of the saturation magnetization. These findings underscore the importance of maintaining high magnetic order and coherence to realize strong photon–magnon coupling.

To further validate the transmission characteristics of the coupled resonator–YIG system, the transmission coefficient $\lvert S_{21} \rvert$ was computed and plotted for the same values using an analytical expression derived from Eq. (7) within the framework of input-output theory \cite{rao2020interactions}, expressed as:
\vspace{-0.8cm}
\begin{equation}
    S_{21} = \Gamma \frac{j^+}{j_0^{+}} = \Gamma \frac{\omega^2 \left( \omega - \omega_r + i\alpha\omega \right)}{det\left( \Omega \right)}
    \label{eq:eleven}
\end{equation}
where, $\Gamma$ is the normalized parameter. Figs. 2(e)–2(h) show the calculated $\lvert S_{21} \rvert$ spectra mapped over the $f-H$ plane. As $M_{\mathrm{s}}$ decreases, the coupling center shifts clearly toward higher magnetic fields in the theoretical results, consistent with the simulation data. This shift arises from the behavior of the magnon resonance frequency, which follows the Kittel equation: $\omega_r = \gamma \sqrt{H\left(H+M_{\mathrm{s}}\right)}$. A reduction in $M_{\mathrm{s}}$ lowers the magnon resonance frequency. Since the HRR photon mode operates at a fixed frequency of 5.33 GHz, resonance requires the magnon mode to match this frequency. To compensate for the reduced magnon frequency due to a lower $M_{\mathrm{s}}$, a higher external magnetic field is required to bring the magnon mode back into resonance with the photon mode. This explains the observed shift of the coupling center toward higher magnetic fields with decreasing magnetization. The strong agreement between theoretical predictions and simulation results further validates the model.

\subsection{\label{sec:citeref} Effect of damping on Photon-Magnon Interactions}
Previous studies on PMC hybrid systems have demonstrated that damping plays a critical role in shaping the anticrossing dispersion and can significantly alter the observed coupling strength \cite{bhoi2019photon,huebl2013high,zhang2014strongly}. In broader contexts such as non-Hermitian and optomechanical systems, it has been shown that the nature of the eigenvalue dispersion whether it exhibits anticrossing or level attraction (i.e., crossing behavior) can depend on the imbalance in dissipation rates between the two interacting subsystems, even for a given coupling strength \cite{harder2018level}.

To systematically explore the influence of magnon damping on the PMC behavior in a strongly coupled system, we extended our CST microwave simulations by varying both the $M_{\mathrm{s}}$ and the $\alpha$ of the YIG film. For each of the previously studied magnetization values (175 mT, 150 mT, 120 mT, and 90 mT), additional simulations were carried out using two elevated damping values: $\alpha = 1.4 \times 10^{-3}$, and $2.1 \times 10^{-2}$. The resulting transmission spectra, displayed in Figs. S2(a)–S2(h) of the supplementary material, reveal that the coupling centers resulting from photon–magnon interactions consistently shift toward higher magnetic fields with decreasing $M_{\mathrm{s}}$, irrespective of the damping level. This trend is consistent with our earlier findings and further confirms the robustness of the theoretical model.

To quantify the effect of damping on coupling strength, the dispersion curves were fitted using Eqs. (9) and (10), as indicated by the black solid lines in Figs. S2(a)–S2(h). The extracted values of the coupling strength and associated constants for each combination of $M_{\mathrm{s}}$ and $\alpha$ are summarized in Table I. Furthermore, a phase diagram of the coupling constant $K$ as a function of magnetization and damping is presented in Fig. S3. Theoretical calculations reveal that when the damping parameter of YIG reaches $\alpha_3 = 2.1 \times 10^{-2}$, the system enters the Purcell regime \cite{zhao2023control,zhang2014strongly}, where the radiative decay of the magnon mode dominates over its intrinsic losses. Interestingly, in this regime, we also observed a spontaneous increase in the damping of the high-quality resonator (HRR), with the damping parameter rising to $\beta = 8.16 \times 10^{-3}$ \cite{verma2025hybrid}. While these results are compelling, a detailed analysis of the emergent Purcell effect and the dynamic interplay of dissipation in this regime lies beyond the scope of the present work and will be addressed in a forthcoming publication.

To visualize the observed trends, Fig. 3(a) plots the coupling strength as a function of damping for all magnetization values, while Fig. 3(b) presents the variation of coupling strength with saturation magnetization for different YIG damping values. These results clearly indicate that the coupling strength in PMC systems decreases both with increasing damping and with decreasing saturation magnetization. This underscores the importance of optimizing both magnetic order and damping characteristics in magnonic materials to achieve and maintain strong photon–magnon coupling.

To gain deeper insights into the interdependence of the $\alpha$ and $M_{\mathrm{s}}$ on the photon–magnon coupling strength $\Delta$, we constructed a phase diagram of the anticrossing behavior in the $\alpha-M_{\mathrm{s}}$ parameter space. This diagram, shown in Fig. 4(a), was generated using Eq. (10), with the coupling constant $K$ values taken from Table I, the resonator damping parameter fixed at $\beta = 4.69 \times 10^{-3}$, and $\omega_c/2\pi$ = 5.33 GHz.  In the phase diagram, the black solid line delineates a contour of constant coupling strength, separating regions of varying $\Delta$. The color gradient visualizes the variation in $\Delta$ across the $\alpha-M_{\mathrm{s}}$ plane: darker red tones signify stronger coupling, while dark blue tones indicate weaker coupling. As expected from theory and simulations, the coupling strength increases with increasing saturation magnetization and decreases with increasing magnon damping. This is consistent with the relation $\Delta \propto \sqrt{M_{\mathrm{s}}}$ and the known detrimental effect of damping on coherent coupling. The dark blue region, located at the intersection of high damping and low magnetization, corresponds to the weakest coupling regime. In contrast, the dark red region characterized by low damping and high magnetization marks the strongest coupling regime.

\begin{figure}[htbp]
    \centering
    \includegraphics[width=\linewidth]{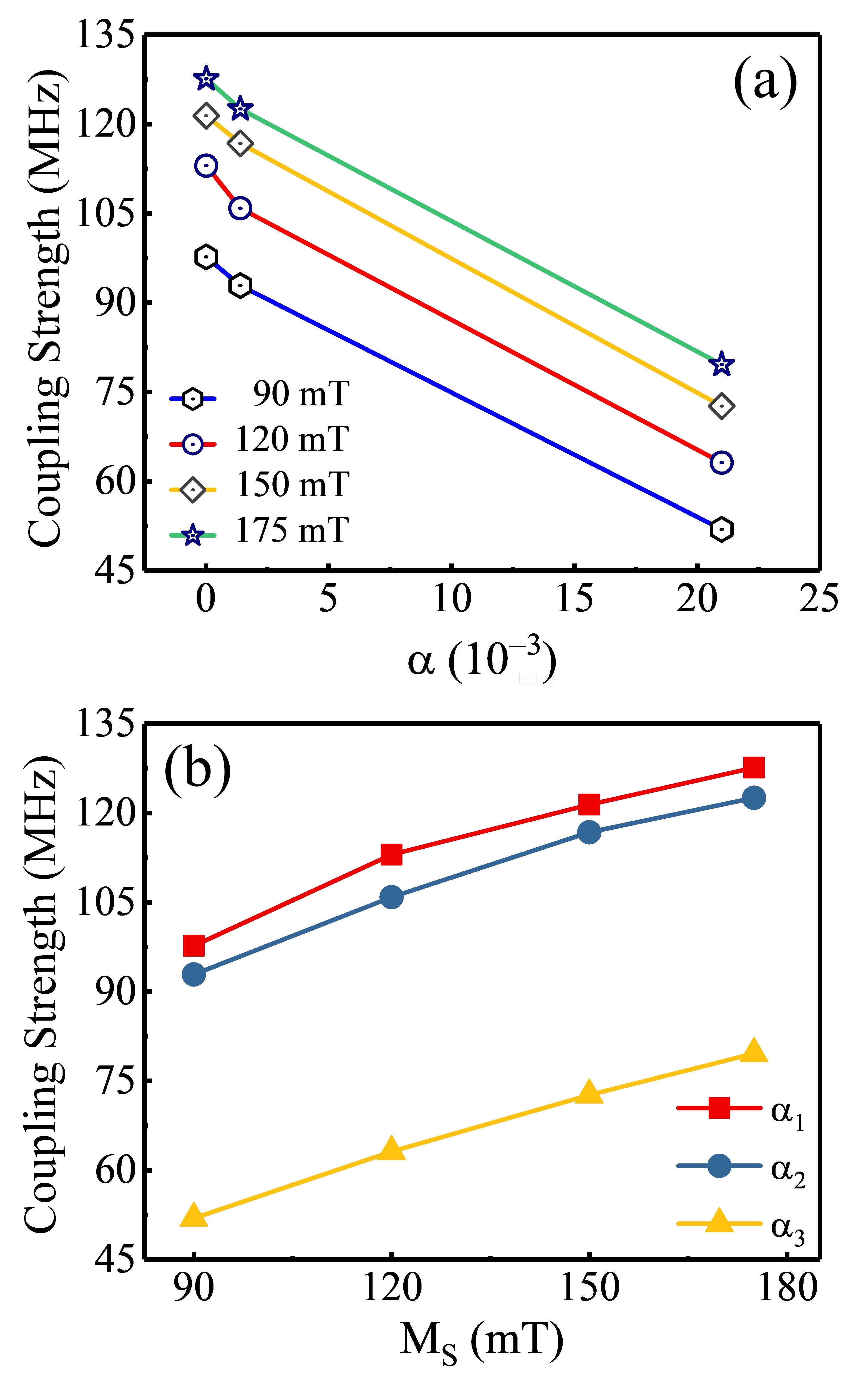}
    \caption{Coupling strength $\left( \Delta \right)$ as a function of (a) magnon damping $\left( \alpha \right)$ for different values of saturation magnetization $M_{\mathrm{s}}$, and (b) saturation magnetization $M_{\mathrm{s}}$ for different values of magnon damping $\left( \alpha \right)$.}
    \label{F3}
\end{figure}

To generalize this behavior for varying resonator damping, Fig. 4(b) shows contours of constant coupling strength for different values of $\beta$ (such as 0.001, 0.00469, 0.008, 0.01, 0.02, 0.03), with their corresponding colormap representations provided in Fig. S4 of the supplementary material. These additional plots illustrate how changes in the HRR’s intrinsic damping influence the coupling landscape. As $\beta$ increases, the contours shift downward in the $\alpha$ axis, indicating that higher HRR damping leads to a reduction in the effective coupling strength at a given magnetization. Interestingly, for larger values of $\beta$, the upper segment of the contour begins to bend rightward toward higher $\alpha$ and intersects with the upper portion of the previous contour. This behavior becomes particularly prominent for $\beta$ = 0.03, where the contour undergoes a significant shape deformation (also highlighted in Fig. S4(g)). In this regime, where the magnon damping is high and magnetization is low, the contours reveal an unexpected result: stronger coupling can still be achieved if the photon mode is embedded in a resonator with sufficiently high intrinsic damping. This suggests a nontrivial interplay between the dissipative properties of both subsystems, hinting at the possibility of entering non-Hermitian or Purcell-like coupling regimes. These findings demonstrate that the coupling strength in PMC systems can be flexibly tuned not only through intrinsic material parameters such as $M_{\mathrm{s}}$ and $\alpha$ but also via the deliberate engineering of resonator losses $\beta$. This multidimensional tuning capability offers a powerful strategy for tailoring hybrid system dynamics.

\begin{figure}[htbp]
    \centering
    \includegraphics[width=\linewidth]{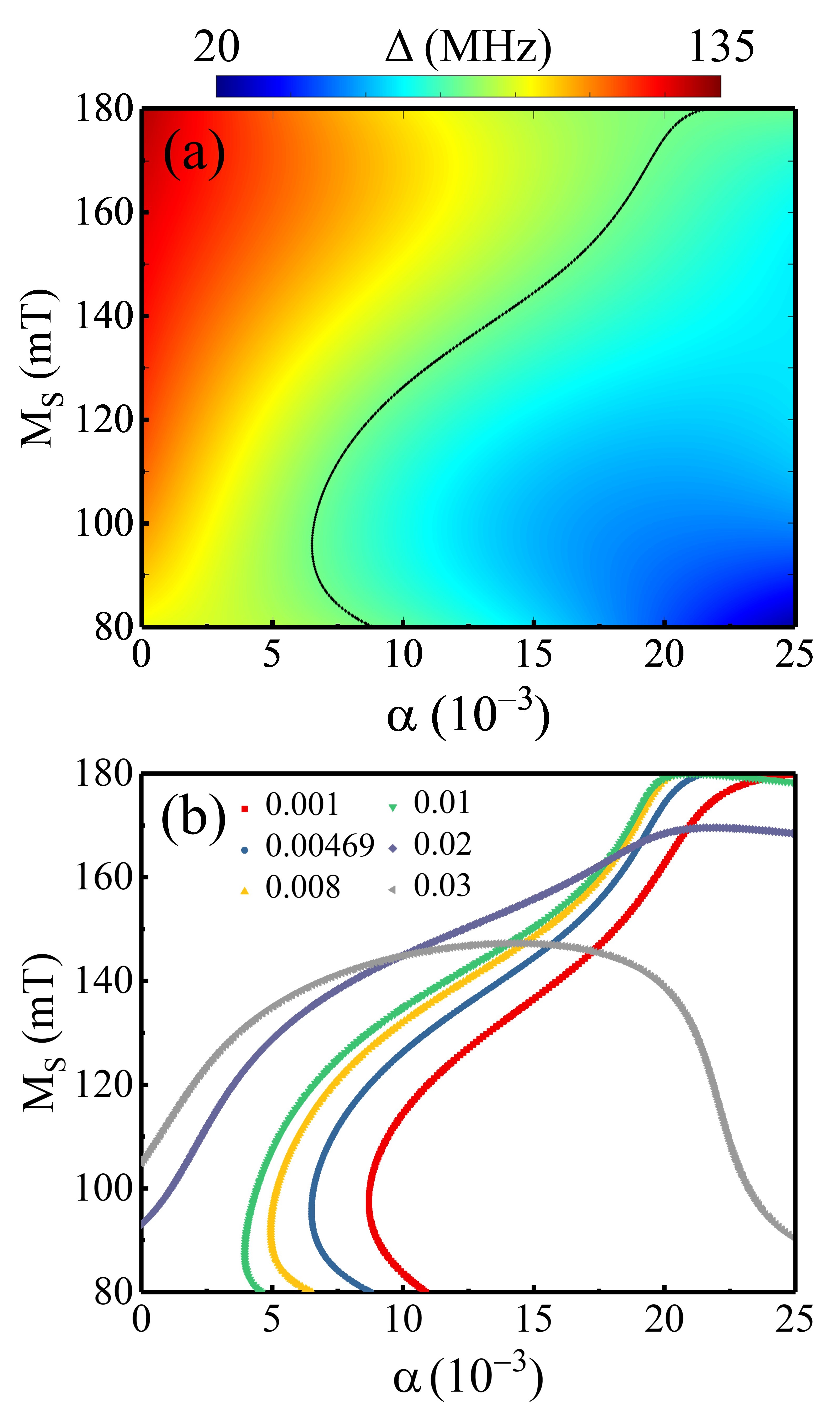}
    \caption{(a) Analytically calculated phase diagram illustrating different types of anti-crossing dispersion on the $\left( M_{\mathrm{s}} - \alpha \right)$ plane for an HRR damping value of $\beta = 4.69 \times 10^{-3}$. The color gradient represents the variation in coupling strength $\left( \Delta \right)$, while the solid black line denotes the contour of constant coupling strength, separating the regions of strong and weak coupling. (b) Contours of constant coupling strength plotted for different values of $\beta$, highlighting the effect of HRR damping on the strong–weak coupling boundary.}
    \label{F4}
\end{figure}

\begin{table*}
\caption{\label{tab:table3}Coupling parameters, coupling strength $\left( \Delta \right)$ and coupling constant $\left( K \right)$, corresponding to different values of saturation magnetization $M_{\mathrm{s}}$.}
\begin{ruledtabular}
\begin{tabular}{ccccccc}
 &\multicolumn{3}{c}{Coupling Strength (MHz)}&\multicolumn{3}{c}{Coupling Constant (K)}\\
 $M_\mathrm{S} (mT)$&$\alpha_1=1.4 \times 10^{-5}$&$\alpha_2=1.4 \times 10^{-3}$&$\alpha_3=2.1 \times 10^{-2}$&$\alpha_1=1.4 \times 10^{-5}$&$\alpha_2=1.4 \times 10^{-3}$&$\alpha_3=2.1 \times 10^{-2}$\\ \hline

 90&97.68&92.88&51.93&0.038&0.036&0.024 \\
 120&113&105.86&63.14&0.038&0.0355&0.024 \\
 150&121.4&116.76&72.63&0.0365&0.035&0.024 \\
 175&127.6&122.55&79.6&0.0355&0.034&0.024 \\
\end{tabular}
\end{ruledtabular}
\end{table*}

\begin{figure}
    \centering
    \includegraphics[width=\linewidth]{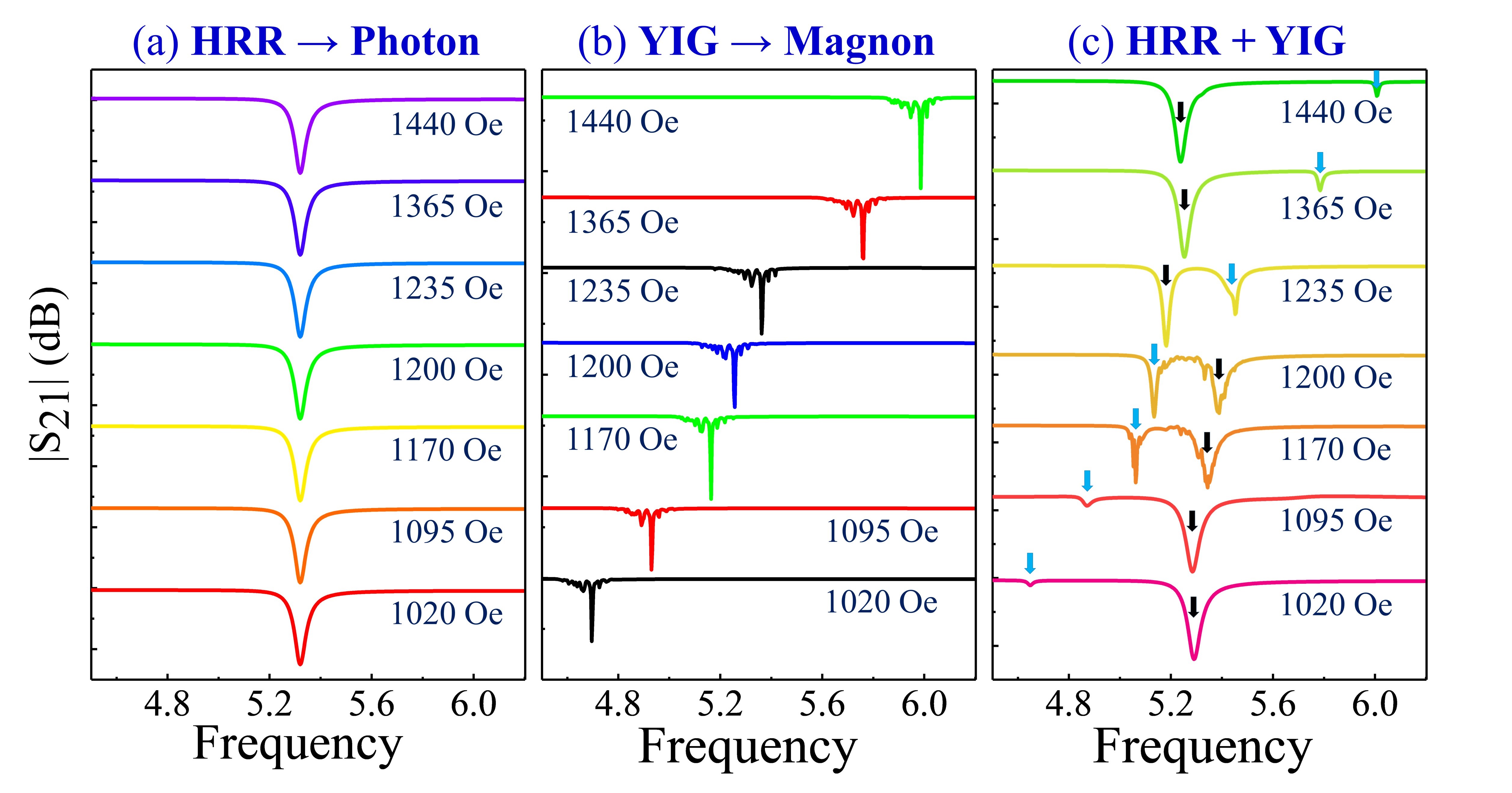}
    \caption{ $\lvert S_{21} \rvert$ spectra as a function of microwave frequency for different magnetic field values applied along the $y$-direction across our hybrid system comprising an HRR and a YIG film: (a) HRR only, (b) YIG film only, and (c) the planar hybrid system of HRR and YIG thin film.}
    \label{FS1}
\end{figure}
\begin{figure}
    \centering
    \includegraphics[width=\linewidth]{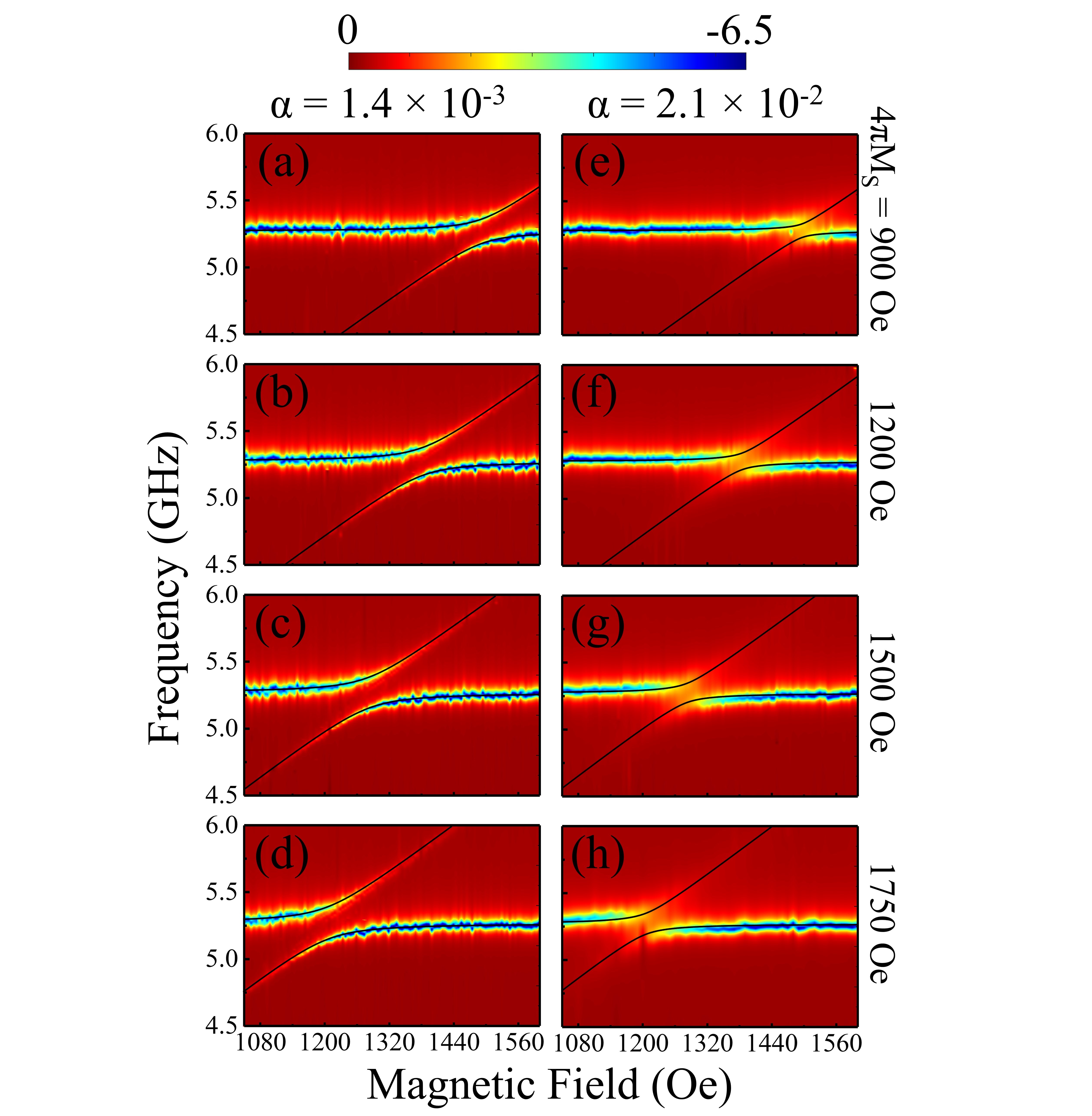}
    \caption{Simulated $\lvert S_{21} \rvert$ power spectra demonstrating the effect of magnetic damping on coupling strength. The spectra are shown for two different damping values, $\alpha_1 = 1.4 \times 10^{-3}$ and $\alpha_2 = 2.1 \times 10^{-2}$, to highlight how increased damping influences the coupling behavior in the hybrid system.}
    \label{FS2}
\end{figure}
\begin{figure}[t]
    \centering
    \includegraphics[width=\linewidth]{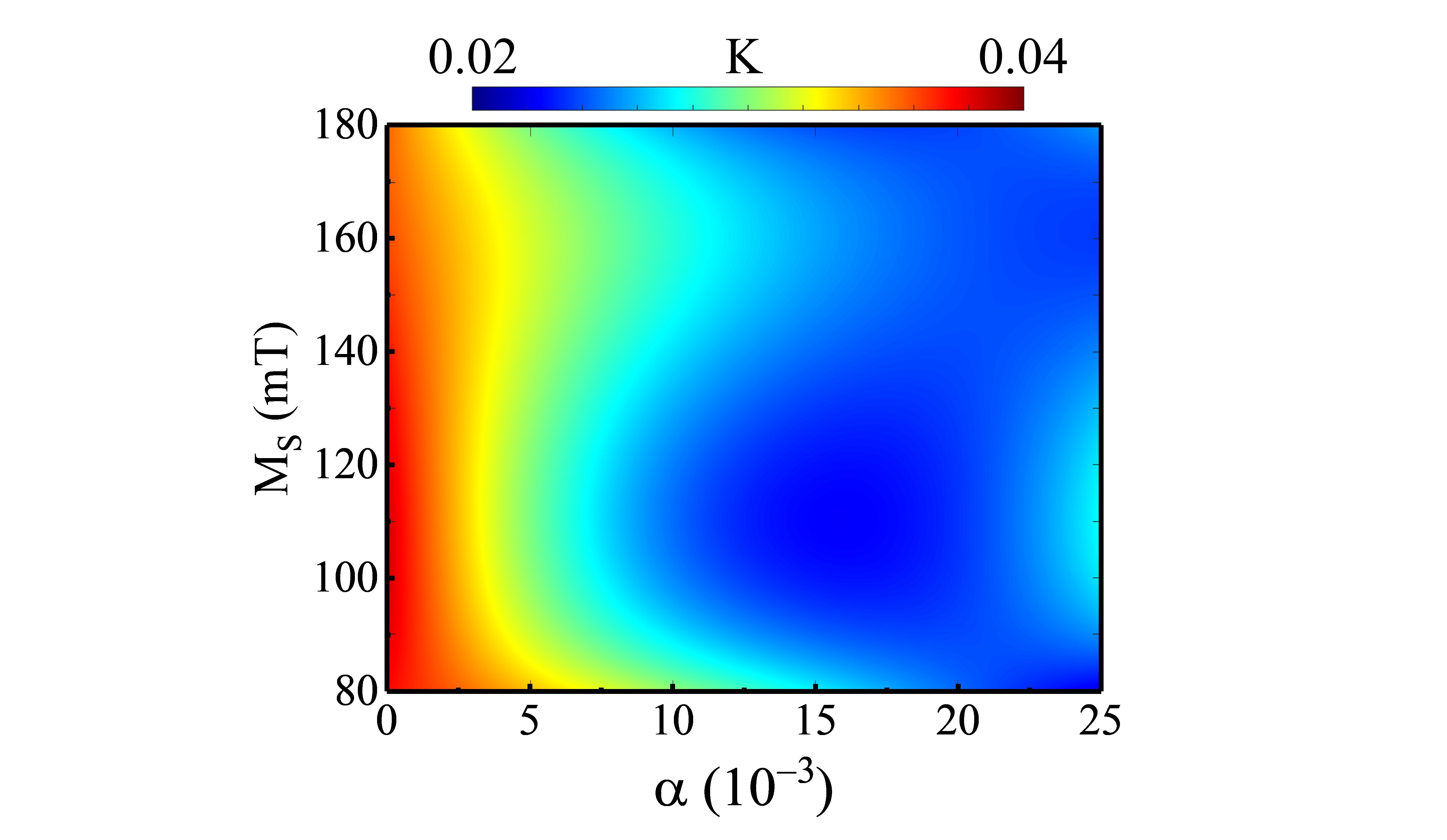}
    \caption{Phase diagram of the coupling constant $K$ as a function of magnetization for different damping values, illustrating the interplay between magnetic properties and coupling strength in the hybrid system}
    \label{FS3}
\end{figure}
\begin{figure}[t]
    \centering
    \includegraphics[width=\linewidth]{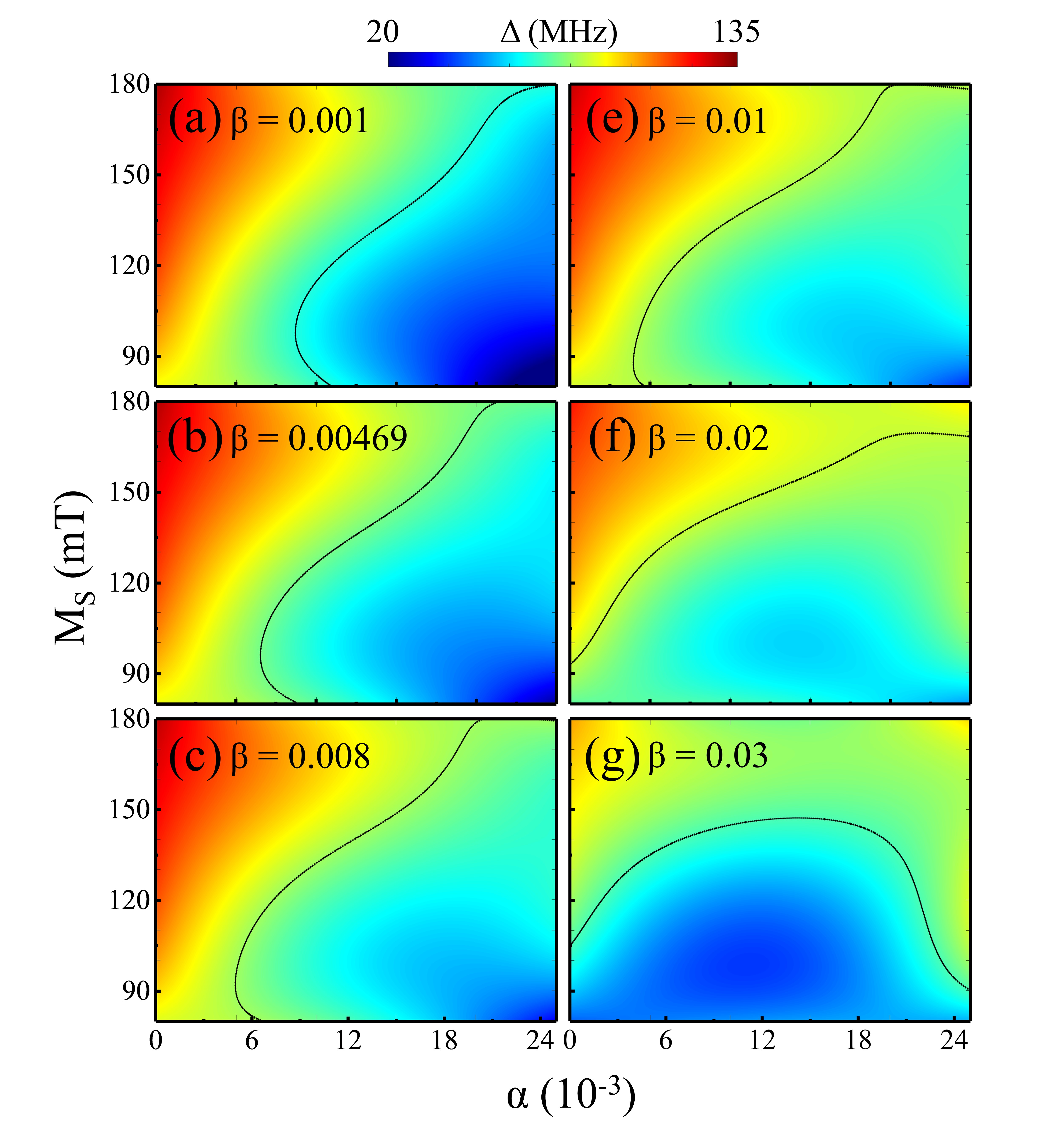}
    \caption{Analytically calculated phase diagrams illustrating the anticrossing behavior in the $\alpha$–$M_S$ parameter space for different values of $\beta$. The diagrams reveal how variations in $\beta$ influence the boundaries and regions of strong and weak coupling in the hybrid system}
    \label{FS4}
\end{figure}

In practical terms, magnetization and damping can be controlled through several well-established methods. The $M_{\mathrm{s}}$ is highly sensitive to factors such as temperature, chemical composition, and structural order. For example, increasing the temperature reduces magnetic order and thus lowers $M_{\mathrm{s}}$, as predicted by Bloch's law for ferromagnets \cite{kittel2018introduction}. Similarly, introducing non-magnetic dopants or changing stoichiometry in YIG films (e.g., substituting Fe with Ga or Al) can reduce the net spin density, thereby tuning $M_{\mathrm{s}}$ \cite{sharma2018magnetic}. On the other hand, the Gilbert damping parameter can be engineered through interface engineering, impurity doping, or varying film thickness. For instance, increasing surface roughness or introducing rare-earth impurities in YIG (e.g., Tb, Dy) has been shown to significantly enhance $\alpha$ by increasing spin-lattice relaxation and magnon-phonon scattering \cite{bhoi2018stress,sharma2018magnetic}. Furthermore, deposition techniques such as pulsed laser deposition (PLD) and liquid-phase epitaxy (LPE) allow precise control over film crystallinity and defect density, thereby offering pathways to tailor both $M_{\mathrm{s}}$ and $\alpha$ \cite{bhoi2013fmr,will2023negligible,sun2013damping}. Therefore, this approach not only provides a new degree of freedom in coupling control but also enables access to previously unexplored regimes of hybrid interactions, such as enhanced non-Hermitian dynamics \cite{wang2019non}, tunable exceptional surfaces \cite{ergoktas2022topological,cerjan2019whole}, and expanded nonreciprocal bandwidths \cite{mann2019nonreciprocal}. Harnessing magnetization-based tuning along with individual damping control of magnons and photons could significantly advance PMC research. It holds particular promise for the development of reconfigurable, chip-scale hybrid quantum systems, where achieving an optimal balance between coherent interaction strength and dissipation is crucial. Such control is vital for next-generation applications in quantum transduction, information storage, and on-chip coherent sensing.

\section{\label{sec:level1}Conclusion}

In conclusion, we present a novel method to tune the coupling strength in photon-magnon hybrid systems by manipulating the magnetic properties of the magnonic material, specifically saturation magnetization, damping, or a combination of both. Our results demonstrate that the coupling strength of HRR-YIG hybridized modes can be effectively controlled through the saturation magnetization of YIG. A decrease in magnetization or an increase in magnon damping leads to a noticeable reduction in coupling strength. These findings are validated using an electrodynamic model, which shows excellent agreement with simulation results. Furthermore, we propose a generalized framework for tailoring the coupling strength across a wide range of system parameters. Such understanding is crucial for the design of next-generation hybrid quantum devices, where robust and tunable photon–magnon interactions are essential for applications in quantum information processing, nonreciprocal microwave components, and coherent signal transduction.
\begin{acknowledgments}
The work was supported by the Council of Science and Technology, Uttar Pradesh (CSTUP), (Project Id: 2470, CST, U.P. sanction No: CST/D-1520). B. Bhoi acknowledges support by the Science and Engineering Research Board (SERB) India- SRG/2023/001355. R. Singh acknowledges support from the Council of Science and Technology, Uttar Pradesh (CSTUP), (Project Id: 4482). S. Verma acknowledges the Ministry of Education, Government of India, for the Prime Minister’s Research Fellowship (PMRF ID-1102628).
\end{acknowledgments}

\section{Supplementary Information}
\section*{S1. Transmission Spectra of the HRR, YIG Film, and their Hybrid Configuration}
\addcontentsline{toc}{section}{S1. Transmission Spectra of the HRR, YIG Film, and their Hybrid Configuration}
\label{sec:S1}

To understand the coupling between the HRR and magnons, we simulated the transmission spectra ($\lvert S_{21} \rvert$) as a function of microwave frequency and external magnetic field at room temperature for three cases: HRR only, YIG only, and the hybrid HRR + YIG system (Supplementary Fig. S1a–c). The HRR shows a constant resonance at 5.33 GHz, independent of magnetic field, with an estimated damping parameter $\beta \approx 4.7 \times 10^{-3}$. In contrast, the YIG film exhibits a resonance peak that shifts with the magnetic field, consistent with the Kittel relation. The hybrid system shows two distinct peaks—one field-independent (photon mode) and one field-dependent (magnon mode)—exhibiting clear anticrossing behavior, indicative of strong magnon-photon coupling.

\section*{S2. $\lvert S_{21} \rvert$ Power Spectra Illustrating the Effect of Damping on Coupling Strength}
\addcontentsline{toc}{section}{S2. $\lvert S_{21} \rvert$ Power Spectra Illustrating the Effect of Damping on Coupling Strength}
\label{sec:S2}

As noted in the main text, we extended our CST microwave simulations by systematically varying the saturation magnetization ($M_S$) and damping constant ($\alpha$) of the YIG film. For each $M_S$ value (175 mT, 150 mT, 120 mT, and 90 mT), additional simulations were performed using elevated damping values of $\alpha = 1.4 \times 10^{-3}$ and $2.1 \times 10^{-2}$ to analyze their influence on the coupling strength.

\section*{S3. Dependence of the Coupling Constant $K$ on Magnetization and Damping Parameters}
\addcontentsline{toc}{section}{S3. Dependence of the Coupling Constant $K$ on Magnetization and Damping Parameters}
\label{sec:S3}
The phase diagram presented illustrates the variation of the coupling constant $K$ as a function of magnetization for different damping values. This diagram highlights the interplay between the magnetic properties, such as saturation magnetization ($M_S$), and the coupling strength in the hybrid system. By varying damping ($\alpha$), the phase diagram showcases how changes in damping influence the coupling behavior, revealing distinct regions where the coupling strength is either enhanced or reduced. This analysis provides key insights into the role of damping in controlling the coupling dynamics in magneto-photonic systems.

\section*{S4. Analytical Phase Diagrams for Different $\beta$ Values}
\addcontentsline{toc}{section}{S4. Analytical Phase Diagrams for Different $\beta$ Values}
\label{sec:S4}
The analytically calculated phase diagrams illustrate the anticrossing behavior in the $\alpha$–$M_S$ parameter space for different values of $\beta$. These diagrams highlight how variations in $\beta$ impact the boundaries between regions of strong and weak coupling in the hybrid system. By modifying $\beta$, the phase diagrams reveal significant shifts in the coupling characteristics, providing a deeper understanding of how damping influences the magnon-photon interaction in the hybrid system.

\nocite{*}

%

\end{document}